%% file: main.tex
\documentclass[10pt,conference]{IEEEtran}
\usepackage[disable]{todonotes}
\usepackage{fancybox,mdframed}
\usepackage{blindtext}
\usepackage{tikz}
\usetikzlibrary{arrows, shapes}
\usetikzlibrary{arrows.meta}
\usetikzlibrary{fit}
\usepackage{booktabs}
\usepackage{multirow}
\usepackage{subcaption}
\usepackage{booktabs}
\usepackage{colortbl}
\usepackage[bookmarks=false]{hyperref}
\usepackage{graphicx}
\usepackage{balance}
\usepackage{lastpage}
\usepackage{float}

\begin{document}
%
\title{A Replication Study on Code Comprehension and Expertise using Lightweight Biometric Sensors}
\author{
\IEEEauthorblockN{Davide Fucci\IEEEauthorrefmark{1}, Daniela Girardi\IEEEauthorrefmark{2}, Nicole Novielli\IEEEauthorrefmark{2}, Luigi Quaranta\IEEEauthorrefmark{2}, Filippo Lanubile\IEEEauthorrefmark{2}}
\IEEEauthorblockA{\IEEEauthorrefmark{1}University of Hamburg, Germany, fucci@informatik.uni-hamburg.de}
\IEEEauthorblockA{\IEEEauthorrefmark{2}University of Bari Aldo Moro, Italy, \{daniela.girardi|nicole.novielli|filippo.lanubile\}@uniba.it,l.quaranta6@studenti.uniba.it}
}

\maketitle

\begin{abstract}
Code comprehension has been recently investigated from physiological and cognitive perspectives using medical imaging devices.
Floyd \textit{et al.} (i.e., the original study) used fMRI to classify the type of comprehension tasks performed by developers and relate their results to their expertise.
We replicate the original study using lightweight biometrics sensors. Our study participants---28 undergrads in computer science---performed comprehension tasks on source code and natural language prose.
We developed machine learning models to automatically identify what kind of tasks developers are working on leveraging their brain-, heart-, and skin-related signals.
The best improvement over the original study performance is achieved using solely the heart signal obtained through a single device (BAC 87\% vs. 79.1\%).
Differently from the original study, we did not observe a correlation between the participants' expertise and the classifier performance ($\tau$ = 0.16, $p$ = 0.31).
Our findings show that lightweight biometric sensors can be used to accurately recognize comprehension tasks opening interesting scenarios for research and practice.
\end{abstract}


\input{sections/introduction.tex}
\input{sections/background.tex}
\input{sections/original.tex}
\input{sections/replication.tex}
\input{sections/analysis.tex}
\input{sections/experiment.tex}
\input{sections/discussion.tex}
\input{sections/conclusion.tex}
\bibliographystyle{IEEEtran}
\balance
\bibliography{./bibtex/references}
\end{document}

%% file: sections/introduction.tex
\section{Introduction}
Developers spend a significant amount of time understanding code as code comprehension is crucial for several software development activities, such as code review~\cite{CLB03,LMC04,RCJ08}. Accordingly, researchers have developed several strategies to study this activity~\cite{NMV11,ARM18}. 

Despite its importance, we only have an initial grasp about the role that human physiological factors~\cite{SKA14} play for code comprehension. 
A first step towards understanding the relationship between code comprehension and the underlying cognitive mechanisms is to study to what extent biometric feedback can help discriminating between code and other comprehension tasks---e.g., natural language comprehension. 
To that end, Floyd et al.~\cite{FSW17} (i.e., the original study replicated in this paper) use functional Magnetic Resonance Imaging (fMRI) to build a classifier able to distinguish these two tasks based on brain activity. 
However, their approach, on top of being expensive (approximately \$500/hour~\cite{FSW17}), limits the ecological validity of the results. 
In this paper, we study to what extent we can replicate the results of the original study using lightweight biometric sensors. 

The decision of using lightweight biometric sensors---i.e., non-intrusive, wearable, and affordable devices---to measure human physiology is supported by the results of recent research demonstrating their potential application to software engineering. 
For example, cognitive-aware IDEs can support developers comprehending code  (e.g., during code-review~\cite{ECN19}) and for foster their productivity by monitoring interruptibility~\cite{ZCM17}.
Fritz et al.~\cite{FBM14} uses a combination of eye-tracker, electrodermal activity, and electroencephalography sensors to measure the difficulty of a development task. 
Fakhoury et al.~\cite{FMA18} used fNIRS (a brain imaging technique which uses sensors connected to a portable headband) to study the effects of poor code readability and lexicon on novice developers' code comprehension. 

In this study, we use \textit{electroencephalogram} (EEG), \textit{electrodermal} (EDA), and \textit{heart-related} sensors to replicate the original study. 
Although such biometrics cannot give a detailed account of a developer's cerebral activity (e.g., activated brain areas) as in the case of fMRI, they can be used to sense variation in cognitive load associated with a cognitive task~\cite{RSI98,GLP18}.  
We developed machine learning models that use features extracted from biometric signals collected from 28 participants comprehending source code and natural language. 
Our approach outperforms the one presented in the original study and achieves its best result using only features from signals acquired using a single wearable device.  

The contributions of this paper are:
\begin{itemize}
   \item an approach for automatic recognition of two comprehension tasks leveraging lightweight biometrical sensors; 
   \item an empirical investigation of which combination of physiological sensors and measurements are most effective at predicting a code comprehension task;
   \item a lab package including to replicate our experiment including data collection material, datasets, scripts, and benchmarks to run and evaluate the machine learning models.\footnote{https://github.com/collab-uniba/Replication\_Package\_ICPC} 
\end{itemize}
This paper is organized according to standard replication report guidelines for software engineering studies~\cite{Car10}.
We consider the current study an external, independent, and differentiated replication~\cite{BCD14,SCV08} as a completely independent set of researchers replicated the original study while making intentional changes to it.

\textbf{Paper organization.} Section~\ref{sec:background} reviews the existing literature regarding the use of biometrics. 
Section~\ref{sec:original} summarizes the original study goals, its settings, and results. 
The main changes to the original study and the details about this replication are reported in Section~\ref{sec:replication} while the machine learning approach is described in Section~\ref{sec:machine_learning}.
Section~\ref{sec:results} reports the results of this replication while Section~\ref{sec:discussion} summarizes its limitation and implications. Finally, we conclude in Section~\ref{sec:conclusion}.

%% file: sections/background.tex
\section{Use of Biometrics in Software Engineering}
\label{sec:background}
The software engineering research community has studied the relationship between the developers' cognitive state---measured using physiological signals---and several aspects of software development, like code comprehension~\cite{PSA18}, productivity~\cite{RHM15}, and software quality~\cite{MF15}. 

Parnin~\cite{Par11} used sub-vocal utterances, emitted by software developers, to study the complexity of two programming task. 
The author used an electromyogram (EMG) to show that those signals are correlated to the cognitive patterns that developers follow when tackling a programming task.
Fritz et al.~\cite{FBM14} combined three physiological features (i.e., eye movement, electrical signal of skin and brain) with a similar goal. 
The authors showed that the three biometrics together provide the  best combination for predicting the difficulty of a task (84.38\% precision, and 69.79\% recall). 
Moreover, they demonstrated that off-the-shelf devices can be used to build accurate, on-line classifiers of difficult code chunks.

Developers’ productivity has recently been the subject of research exploiting physiological measures.
For example, Radevski et al.~\cite{RHM15} proposed a framework for continuous monitoring of developers' productivity based on brain electrical activities.  
M{\"u}ller and Fritz~\cite{MF15} used an ensemble of biometrics to measure the progress and interruptibility of developers performing small development tasks. 
They demonstrated that it is possible to classify the emotions experienced by developers while working on a programming task using biometrics (i.e., brainwave frequency, pupil size, heart rate) with an accuracy of 71\%. 
The progress experienced by developers was predicted at a similar rate, but using a different set of biometrics (i.e., EDA signal, skin temperature, brainwave frequency, and the pupil size).

M{\"u}ller and Fritz~\cite{MF16} investigated the use of physiological measures for real-time identification of code quality concerns in a real-world setting. 
Using heart-, skin-, and brain-related biometrics, the authors identified difficult parts of the system---e.g., low-quality code containing bugs. 
The authors provide some evidence that biometrics can outperform traditional code-related metrics to identify quality issues in a code base. 

\textbf{For Code Comprehension.} As far as code comprehension is concerned, two similar studies,  Siegmund et al.~\cite{SKA14} and Ikutani and Uwano~\cite{IU14}, assessed the brain activity of developers involved in code comprehension tasks.
Siegmund et al.~\cite{SKA14} used fMRI to show clear activation patterns in five regions of the brain all related to language processing, working memory, and attention.
The study by Ikutani and Uwano~\cite{IU14} uses near-infrared spectroscopy to show that different parts of the brain are activated during code comprehension with respect to a specific sub-task. For example, they distinguish between the areas activated by the workload necessary to memorize a variable and the ones activated by arithmetic calculation.

More recently, Peitek et al.~\cite{PSA18} used fMRI to monitor the brain activity of 28 participants involved in the comprehension of 12 source code snippets.
Their results show that distinct areas of the brain are activated during such a task.
Moreover, the activation patterns suggest that natural language processing is essential for code comprehension.
To get a more comprehensive view of the strategies adopted by developers when comprehending source code, Peitek et al.~\cite{PSP18} obtained simultaneous measurements of fMRI and eye-tracking devices. 
They showed strong activation of specific brain areas when code beacons are available.
However, their setup was subject to data loss---complete fMRI and eye-tracking data could be collected for 10 out of the 22 participants. 

%% file: sections/original.tex
\section{Original Study}
\label{sec:original}
This section summarizes the original study, giving an overview of its settings, methodology, and results. 
\subsection{Research Question}
The original study explored the use of fMRI and machine learning techniques to automatically distinguish between code comprehension, code review, and prose review tasks. 
Moreover, it investigated whether the neural representation of programming and natural languages changes depending on the developer's expertise.

To guide their research, the authors formulated the following research questions~\cite{FSW17}:
\begin{itemize}
    \item \textit{RQ1} - Can we classify which task a participant is undertaking based on patterns of brain activation?
    \item \textit{RQ2} - Can we relate tasks to brain regions?
    \item \textit{RQ3} - Can we relate expertise to classification accuracy?
\end{itemize}

\subsection{Participants and Context}
The original study involved 29 students (18 men, 11 women) at the University of Virginia (USA) with basic experience in the C programming language. 
Among them, two were computer science graduate students, nine were undergraduates in the College of Arts and Sciences, and 18 were undergraduates in the College of Engineering.
All participants were right-handed native English speakers, had normal or corrected-to-normal vision, and reported no history of neuropsychological disorders.
The authors rewarded the students for their participation with monetary compensation and extra university credits.
The experiment was conducted at the University of Virginia (USA).
\subsection{Artifacts}
The authors prepared three types of artifacts according to the experimental tasks.
\begin{enumerate}
    \item Code snippets with related software maintenance questions. 
    \item  Patches from GitHub Pull Request including code diff and comments. 
    \item  English texts with simple editing request markup. 
\end{enumerate}
The artifacts are available at the original experiment website.\footnote{\url{https://web.eecs.umich.edu/~weimerw/fmri.html}}

\subsection{Design}
\label{sec:original_design}
The experiment consisted of three tasks, code comprehension, code review, and prose review. 
The tasks were presented as visual stimuli on a special screen installed in the fMRI scanner. 
Before beginning the experiment, participants signed a consent form for personal data treatment. 
The original experiment started by showing the participants an instructional video explaining the goal and the different steps of the study.
The participants entered the fMRI scanner for an initial anatomical scan.
Then, they performed the experimental tasks consisting of four 11-minute sessions where blocks of code review, code comprehension, and prose review were presented in a quasi-random order.
Code comprehension and code review blocks contained three tasks each, whereas prose review blocks were composed of six tasks. 
Tasks containing source code were displayed for 60 seconds and the ones containing prose for 30 seconds.
The participants provided their answer (i.e., accept or reject) through an fMRI-compatible button. 
They were encouraged to respond as quickly and accurately as possible within the allotted time. 
Between tasks, the screen displayed a fixation cross for an random interval between two and eight seconds.
The sessions were completed without interruptions. 

\subsection{Summary of Results}
The original study authors found that neural representations of programming languages and natural language are distinct. 
Specifically, they used Gaussian Process Classification to distinguish between code and prose tasks, achieving a balanced accuracy (BAC) of 79\%. 
They show that neural activity in the prefrontal regions strongly drives this distinction. 
However, their approach performance was lower (BAC = 62\%) when comparing code comprehensions to code review, revealing that these tasks are less distinguishable at a neural level. 

Finally, authors showed a negative correlation between their classifier performance and the participants' expertise ($r$ = 0.44, $p$ = 0.16), indicating that for experts the neural representation of source code and prose are similar. 

%% file: sections/replication.tex
\section{Our Study}
\label{sec:replication}
This section summarizes our replication.
\subsection{Motivation for conducting the replication}
We conducted this replication to broaden the original study results by replacing the observed signal (i.e., neural activity sensed through fMRI) with a different set of signals capturing the same construct (i.e., cognitive effort).
Moreover, we want to increase the ecological validity of the original study by using sensing devices which can be used in real-world settings.

\subsection{Level of interaction with original experimenters}
The authors of the original study did not take part in the replication process; therefore, this replication is to be considered external~\cite{BCD14}.

We reused a subset of 18 source code snippets that the authors of the original study made available in their replication package.

\subsection{Changes to the original experiment}
\input{tables/comparison.tex}
This replication makes explicit changes to the original study.
\begin{enumerate}
    \item Adaptation of the research question;
    \item Partial modification of task presented to the participants through visual stimuli;
    \item Different physiological signals captured from the participants performing the task;
    \item Modifications to the experimental protocol;
    \item Additional machine learning settings.
\end{enumerate}
Table~\ref{tab:comparison} compares the original study settings to this replication.

\textbf{Research Questions.}
In our study, we answer the following research questions:
\begin{itemize}
    \item $RQ_{Clf}$ - Can we classify which task a participant is undertaking based on \textit{signals collected from lightweight biometric sensors?}
    \item $RQ_{Exp}$ - Can we relate expertise to classification accuracy?
\end{itemize}

Our research questions are adapted from those addressed in the original study.
Specifically, $RQ_{Clf}$ is adapted from RQ1 of the original study, which we modify by considering lightweight biometric sensors instead of fMRI.

In the original study, RQ2 investigates whether the tasks are associated with the activation of specific brain areas.
In our study, it is not possible to address this question.
The lightweight EEG device we use in our replication to obtain brain-related signals is not capable of registering activation of brain areas as it allows to only collect the signal from the frontal part of the brain.
Therefore, we decided to discard the original study RQ2 from our replication.
Finally, we address RQ3 from the original study considering the accuracy of each participant best classifier trained using biometrical signals.

\begin{figure*}[!htb]
\centering
\begin{subfigure}[b]{0.440\textwidth}
  \includegraphics[width=\linewidth]{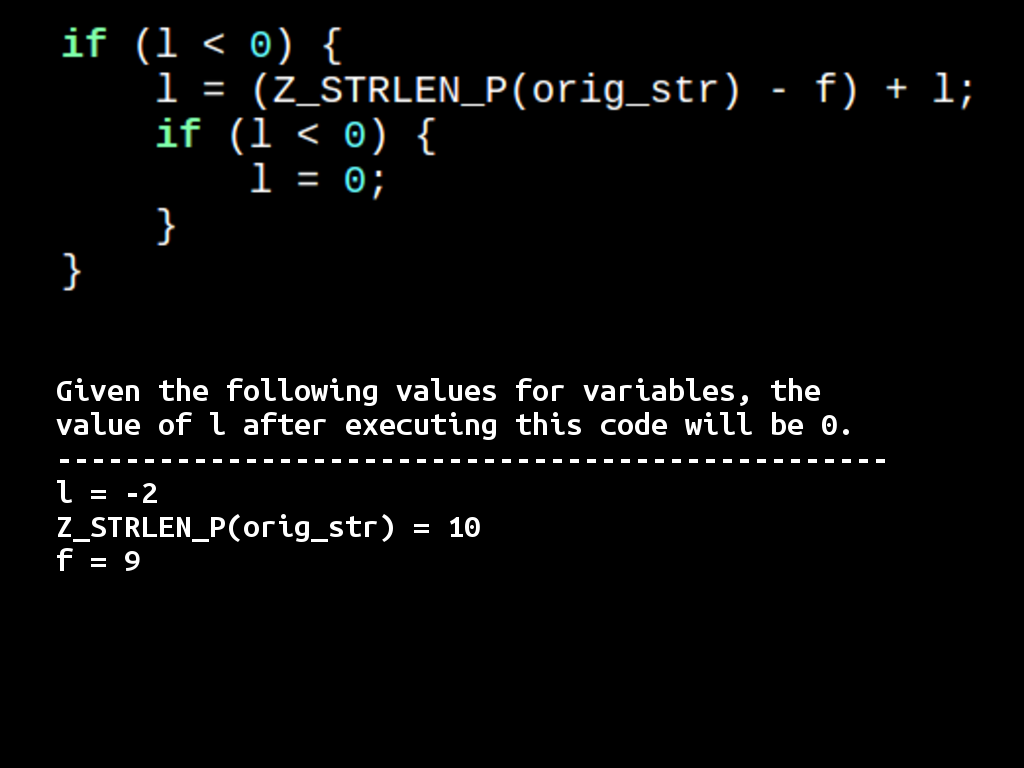}
  \caption{Example of a code snippet used for the code comprehension task.}\label{fig:code_comp}
\end{subfigure}
\begin{subfigure}[b]{0.440\textwidth}
\vspace{0.4cm}
  \includegraphics[width=\linewidth]{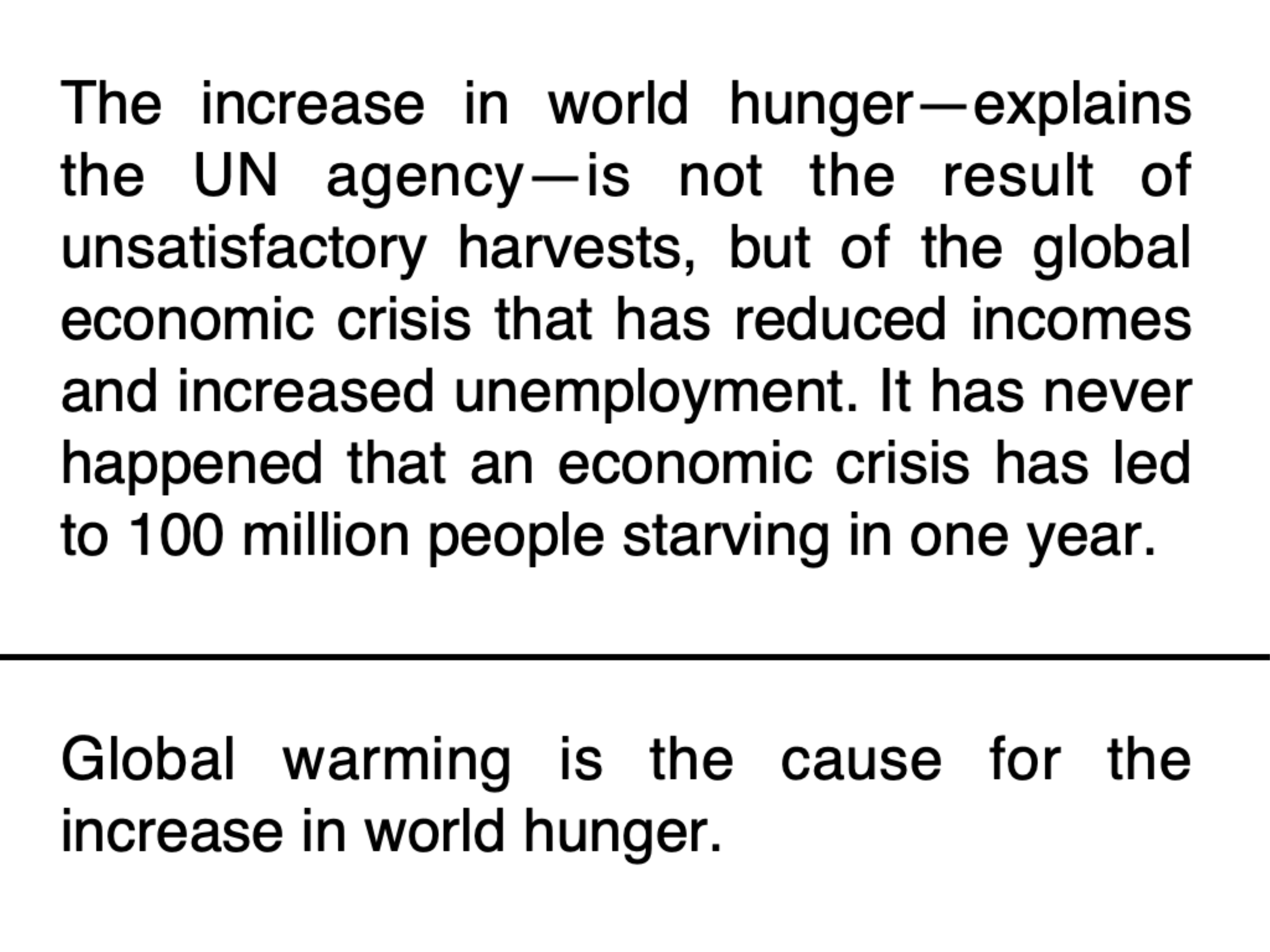}
  \caption{Example of a prose snippet used for the prose comprehension task (translated from Italian to English).}\label{fig:prose_compw}
\end{subfigure}
\caption{Examples of tasks for code and prose comprehension used in this study.}
\end{figure*}

\textbf{Tasks.} In our study, the participants are required to solve a series of code comprehension tasks (see Fig. \ref{fig:code_comp} ) and a series of prose comprehension task (see Fig. \ref{fig:prose_compw}).
As opposed to the original study, we decided to focus only on comprehension; thus, excluding code review and prose review tasks.

Initially, we conducted a pilot study to validate the feasibility of all the original study tasks, including code review.
The participants involved in the pilot (i.e., a Ph.D. student and a researcher in Computer Science) perceived code review tasks as too difficult and the entire experiment as too demanding given the allotted time.
The pilot participants reported that they felt overwhelmed when performing the code review task and that they ended up providing random answers without actually trying to solve the task. Such behavior is a threat to the study validity which led us to discard  the code review tasks.

Finally, to be consistent with the type of activities to compare, we replaced the prose review appearing in the original study with a new prose comprehension task.
We operationalized prose comprehension using standard evaluation exercises for high school students (see Fig. \ref{fig:prose_compw}).
We repeated the pilot study with the same participants, who agreed with the changes.

\textbf{Physiological signals.}
In the original study, the authors used images captured from functional magnetic resonance (fMRI) to build a classifier able to distinguish between the tasks a participant is performing.
The fMRI provides indirect estimates of brain activity by measuring metabolic changes in blood flow and oxygen consumption.
Although this technique allows to understand how the human brain processes software engineering tasks, it is expensive and cannot be used in real-world settings---i.e., to monitor a developer's cognitive activity during daily programming tasks.

\begin{figure}[!hb]
\centering
\begin{subfigure}[b]{0.22\textwidth}
  \includegraphics[width=\columnwidth, scale=0.8]{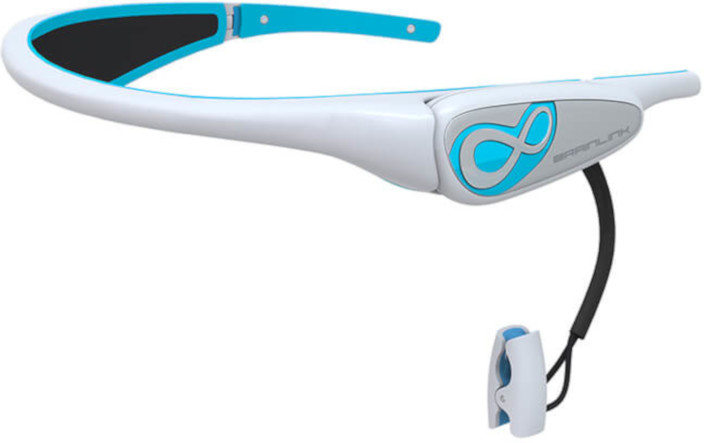}
  \caption{BrainLink EEG headset.}\label{fig:Brainlink}
\end{subfigure}
\qquad
\begin{subfigure}[b]{0.15\textwidth}
  \includegraphics[width=\columnwidth, scale=0.8]{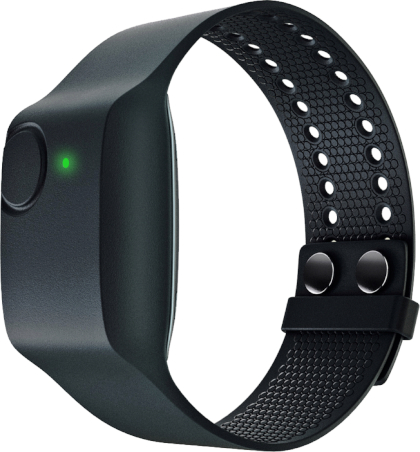}
  \caption{Empatica E4 wristband.}\label{fig:Empatica E4}
\end{subfigure}
\caption{Devices used to measure biometric signals in this study.}\label{fig:Biometrics}
\end{figure}

Thus, we decided to measure other physiological signals which can be recorded using low cost, lightweight biometric sensors.
In our study, we use the BrainLink headset (Fig.~\ref{fig:Brainlink}) to record the electrical activity of the brain (EEG), and the Empatica E4 wristband (Fig.~\ref{fig:Empatica E4}) to record the electrodermal activity of the skin (EDA) and the blood volume pulse (BVP).
The EEG sensor records the electrical activity of the brain through one electrode placed on the surface of the scalp.
The cerebral waves can be categorized based on frequency as delta ($<$4Hz), theta (4-7,5Hz), alpha (4-12,5Hz), beta (13-30Hz), and gamma ($>$ 30Hz).
Delta waves are mainly recorded during sleep, theta waves indicate a decrease of vigilance level, alpha waves are recorded during relaxing moments, beta waves are observed during mental activity demanding attention or concentration, and gamma waves are related to cognitive processes. In addition to raw data for the EEG signal, the BrainLink device extracts metrics related to meditation and attention levels.~\footnote{\url{http://developer.neurosky.com/docs/doku.php?id=esenses_tm}}
The EEG samples the signal at 512Hz.

EDA is constituted by a tonic component, indicating the level of electrical conductivity of the skin (SCL), and a phasic component, representing the phasic changes in electrical conductivity or skin conductance response (SCR)~\cite{BWJR15}.
The device samples the EDA signal at a frequency of  4Hz.

BVP is the volume of blood that passes through tissues in a localized area with each heartbeat.
It is used to calculate the heart rate (HR) and the variation in the time interval between heartbeats or heart rate variability (HRV).
The device samples the BVP signal at a frequency of  64Hz.

\textbf{Experimental protocol.}
We organized the experiment according to the following phases.

\textit{Pre-experimental briefing}. The participant gets acquainted with the settings---e.g., sitting in a comfortable position, adjusting the monitor height.
The experimenter summarizes the upcoming steps and explains how to perform the task.
Subsequently, the participants signs the consent form for personal data treatment.

\begin{figure}
    \centering
    \includegraphics[width=\linewidth]{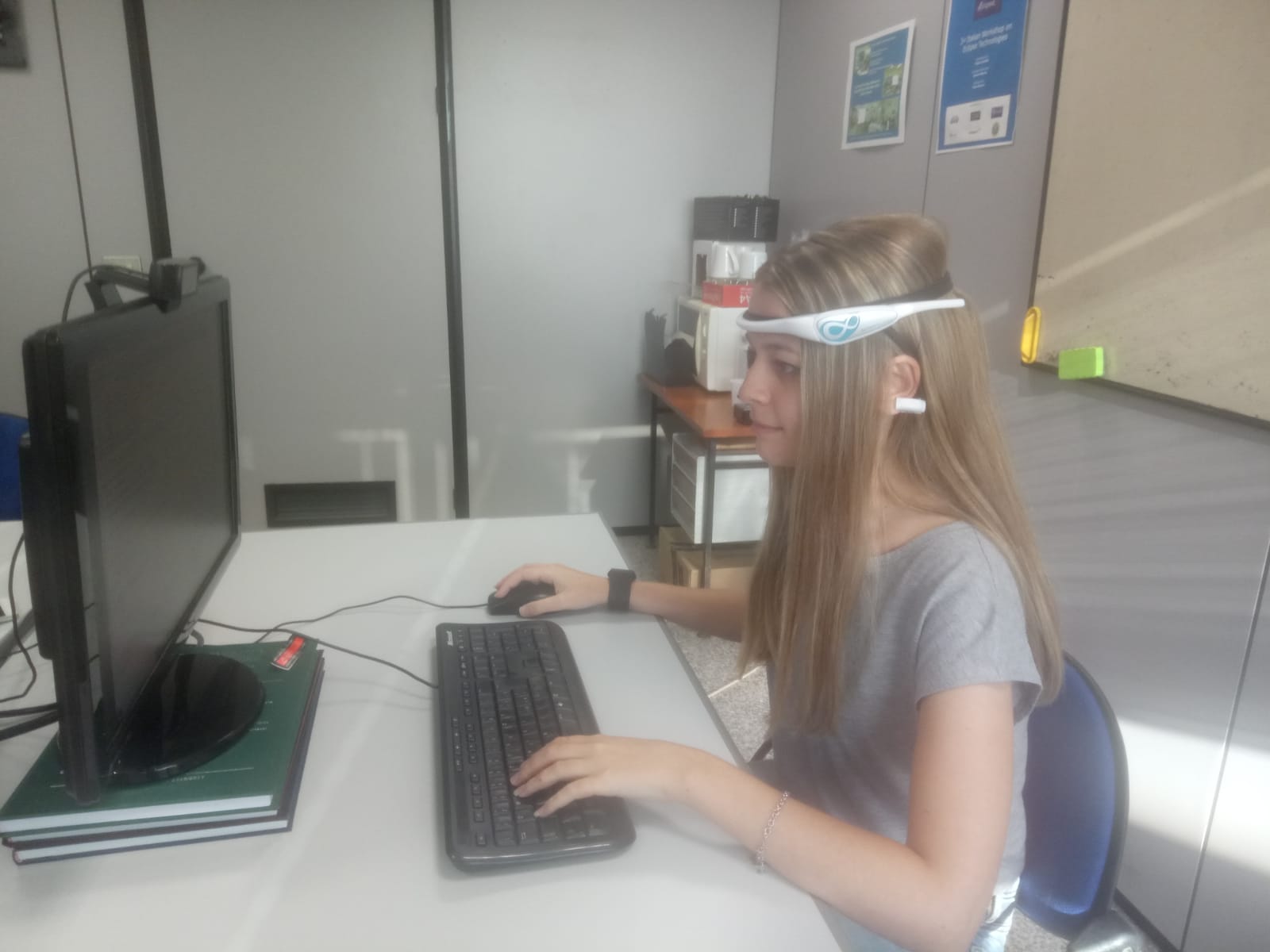}
    \caption{A participant wearing the wristband and headband during the experimental session.}
    \label{fig:exp_session}
\end{figure}

\textit{Device calibration}. The participant wears the biometric sensors (see Figure~\ref{fig:exp_session}) and watches a two-minute fish thank video to collect physiological baselines.\footnote{Sensors 101 workshop. \url{https://github.com/BioStack/Sensors101}}

\textit{Task execution.} The participants performed 27 tasks, divided into three sessions.
The tasks are displayed on a 24-inches monitor connected to a standard desktop computer.
Answers are recorded by pressing the arrow keys (i.e., left arrow to accept and right arrow to reject).
After each session, a 10-second fixation cross is displayed.
Each session is composed of three unique code comprehension and six unique prose comprehension tasks randomly displayed for 60 and 30 seconds respectively.
The total duration of the experiment for one participant was 30 minutes.

\textit{Post-experimental briefing}. The participants can ask questions and give feedback about the experiment.
Finally, they were rewarded with a voucher for a meal.

\textbf{Machine Learning settings}.
We perform the classification using eight different machine learning algorithms whereas only one, Gaussian Process Classifier, was used in the original study.
Moreover, we present two different validation settings, leave-one-run-out cross-validation (LORO-CV, as in the original study) and the additional Hold-out validation.

\subsection{Experimental Sample}
Our goal is to have a sample size comparable to the original study.
Accordingly, we recruited 32 (28 males, four females) Bachelors' students from the Department of Computer Science.
We applied a quota sampling strategy based on students expertise, measured through the number of credits obtained in courses where the C programming language (i.e., the language used for the code comprehension tasks) was used.
At the time of the experiment, the average GPA (grade point average) of the students was 3.0 ($\pm 0.25$).

\textbf{Outliers and dropouts}. Once the experiment was completed, but before analyzing the data, we discarded two participants because they failed to complete more than 30\% of the tasks (e.g., they did not provide an answer within the time allotted).
We interpreted this as a sign of inability or negligence in carrying out the experimental tasks.
Moreover, due to a technical issue with the devices, we discarded two more participants.
Therefore, we considered a total of 28 participants (24 males, four females) during data analysis.

%% file: tables/comparison.tex
\begin{table}[t]
\caption{Settings comparison between the original study and this replication.}
\label{tab:comparison}
\scriptsize
\begin{tabular}{lll}
\toprule
\multicolumn{1}{c}{}                                   & \multicolumn{2}{c}{\textbf{Study}}                                                                                                                                                                  \\ \cmidrule(l){2-3} 
\multicolumn{1}{c}{\multirow{-2}{*}{\textbf{Setting}}} & \multicolumn{1}{c}{\textbf{Original study}}                                             & \multicolumn{1}{c}{\textbf{This replication}}                                                             \\ \midrule
Experiment site                                        & Univ. of Virginia (USA)                                                                 & Univ. of Bari (Italy)                                                                                     \\
\rowcolor[HTML]{EFEFEF} 
\# Participants                                        & 29                                                                                      & 28                                                                                                        \\
Participants experience                                &                 Grads and undergrads                                                                        &            Undergrads                                                                                               \\
\rowcolor[HTML]{EFEFEF} 
\# Task                                                & \begin{tabular}[c]{@{}l@{}}36 tasks \\ four 11-minute sessions\end{tabular}            & \begin{tabular}[c]{@{}l@{}}27 tasks \\ three 6-minute sessions\end{tabular}                              \\
Task type                                              & \begin{tabular}[c]{@{}l@{}}Code comprehension\\ Code review\\ Prose review\end{tabular} & \begin{tabular}[c]{@{}l@{}}Code comprehension\\ Prose comprehension\end{tabular}                          \\
\rowcolor[HTML]{EFEFEF} 
Physiological signal & Neural & \begin{tabular}[c]{@{}l@{}}Neural\\ Skin\\ Heart\end{tabular} \\
Physiologial measure & BOLD &\begin{tabular}[c]{@{}l@{}}EEG\\ EDA\\ BVP, HR, HRV\end{tabular} \\
\rowcolor[HTML]{EFEFEF} 
Device                                                 & fMRI scanner                                                                            & \begin{tabular}[c]{@{}l@{}}BrainLink headset\\ Empatica wristband\end{tabular}                            \\
Classifier                                             & Gaussian Process                                                                        & \begin{tabular}[c]{@{}l@{}}Machine Learning \\(8 algorithms)\end{tabular}                                                                               \\
\rowcolor[HTML]{EFEFEF} 
Classifier validation                                  & LORO-CV                                                                                 & \begin{tabular}[c]{@{}l@{}}LORO-CV\\ Hold-out\end{tabular}                                                \\
Classifier metric                             & Balanced accuracy (BAC)                                                                 & Balanced accuracy (BAC)                                                                                   \\ \bottomrule
\end{tabular}
\end{table}

%% file: sections/analysis.tex
\section{Machine Learning}
\begin{figure*}[h!]
    \centering
    \includegraphics[width=.9\linewidth]{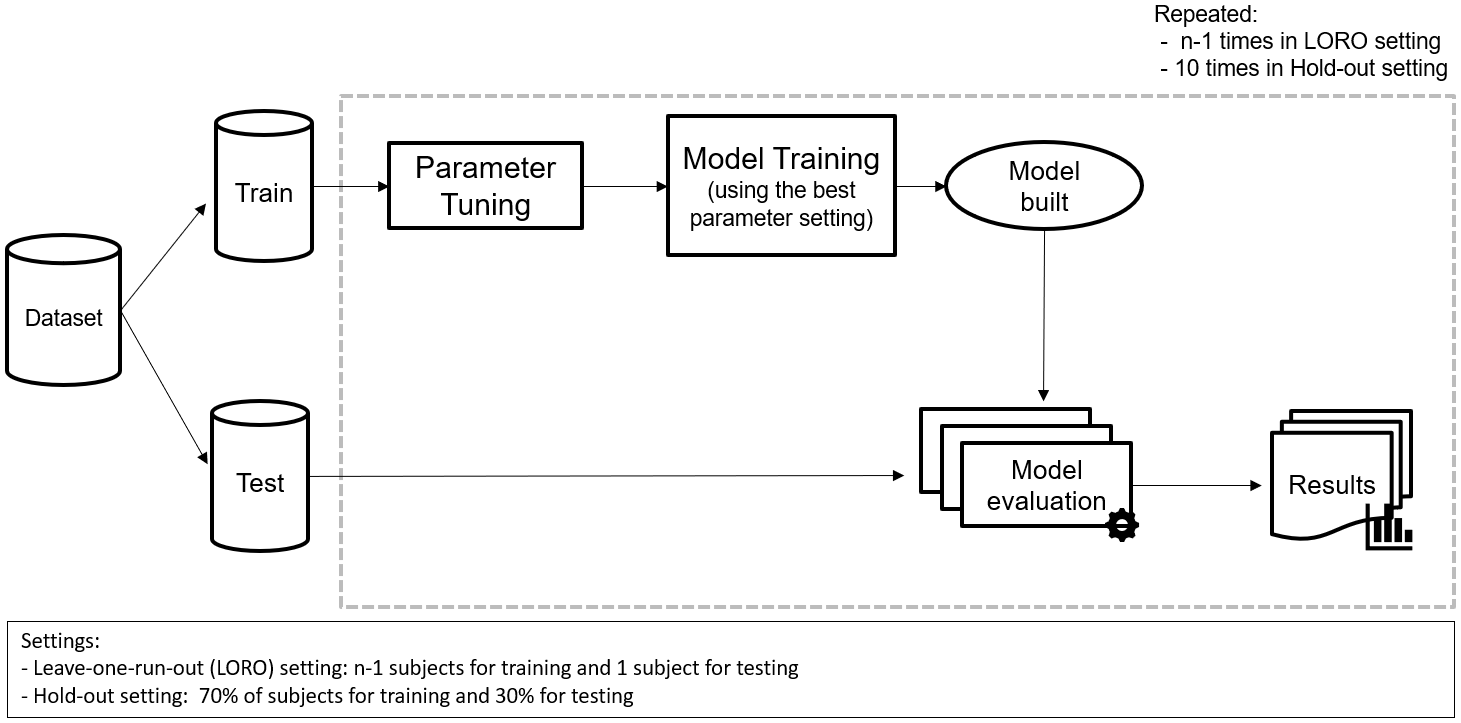}
    \caption{Machine learning pipeline implemented in this study. The evaluation settings include LORO and Hold-out cross-validation.}
    \label{fig:setting}
\end{figure*}

In this section, we report the machine learning approach used to classify the comprehension tasks. The machine learning pipeline implemented in this study is reported in Fig.\ref{fig:setting}.
\label{sec:machine_learning}
\subsection{Dataset}
Each of the 28 participants performed a total of 27 comprehension tasks (i.e., nine code, 18 prose).
They had the possibility of not answering a task question if they did not feel confident.
Therefore, out of the total 756 tasks, we consider the biometric signals associated with the 695 completed ones---469 of prose comprehension and 226 of code comprehension.

\subsection{Preprocessing and Features extraction}
The biometric signals were recorded during the entire experimental session for all the participants.
However, to address our research questions, we only consider the signals associated with the stimuli of interest---i.e., the signals collected between the time a task appeared on the screen and the time the participant provided an answer.
We did not consider signals collected when a participant was not focusing on a task---i.e., when a fixation point was displayed on the screen.
To synchronize the measurement of the biometric signals with the tasks, we applied the following procedure.
\begin{enumerate}
    \item We registered the timestamp at the start of the experiment  \texttt{(t\_start\_experiment});
    \item We saved the name of the task, its type (code or prose), and the timestamp of the answer \texttt{(t\_answer)};
    \item We calculated the timestamp for the start of each task (\texttt{t\_start}) using \texttt{t\_start\_experiment} and the duration associated with each type (60 seconds for code and 30 seconds for prose);
    \item From each biometric signal, we selected the samples recorded between \texttt{t\_start} and \texttt{t\_answer}.
\end{enumerate}

For each participant, we normalize the signals to her baseline using Z$-$score normalization \cite{MF15}.
As it is customary for this kind of studies~\cite{FBM14}, the baseline was calculated considering the last 30 seconds of the fish thank video showed to the participant during the device calibration.

\begin{figure}[!hb]
\begin{tikzpicture}
[> = latex', auto,
block/.style ={rectangle,draw=black, thick,
align=flush center, rounded corners,
minimum height=3em}, ]
\matrix [column sep=5mm,row sep=3mm]
{

\node [block, text width=15mm] (b1) {EEG};&
&\node [block, text width=50mm] (b2) {Band pass filter: \\
$\bullet$ Delta: 0$-$4HZ\\
$\bullet$ Theta: 4$-$8HZ \\
$\bullet$ Alpha: 8$-$12HZ \\
$\bullet$ Beta: 12$-$30HZ \\
$\bullet$ Gamma: $>$30HZ \\

};\\

\node [block, text width=15mm] (b3) {EDA};&
&\node [block, text width=50mm] (b4) {cvxEDA algorithm: \\
$\bullet$ Tonic component\\
$\bullet$ Phasic component \\
};\\

\node [block, text width=15mm] (b5) {BVP};&
&\node [block, text width=50mm] (b6) {Band pass filter:
1$-$8HZ

};\\
};

\draw[->] (b1) -- (b2);
\draw[->] (b3) -- (b4);
\draw[->] (b5) -- (b6);

\end{tikzpicture}
\caption{Filtering strategies for the biometric signals (EEG, EDA, and BVP) collected in this study.}
\label{fig:preprocessing}
\end{figure}
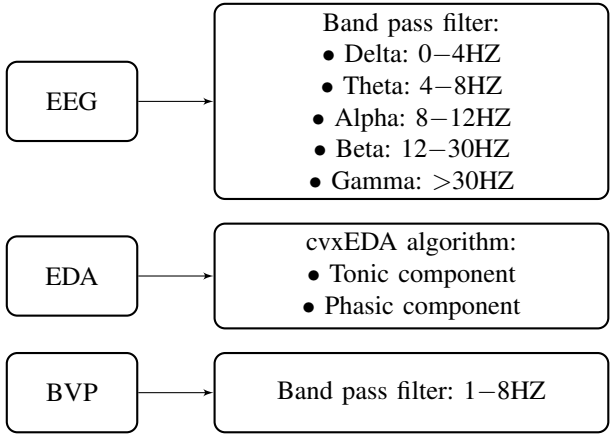

Finally, to maximize the signal information and reduce noise caused by movement, we applied multiple filtering techniques, as reported in Figure~\ref{fig:preprocessing}.

Regarding EEG and BVP, we extract the different frequency bands using a band$-$pass filter at different intervals~\cite{CFSGL11}.
Concerning EDA, we applied the cvxEDA algorithm~\cite{GVLSC16} to extract the tonic and the phasic components.

After signals preprocessing, we extracted the feature presented in Table~\ref{tab:features}.
We selected features based on previous studies~\cite{MF15} in which the same signals were used to train machine learning classifiers for recognizing affective and cognitive states of software developers.
\input{tables/table_features.tex}

In 21 cases, values were missing for HRV features due to noise in the recorded signals.
As suggested in \cite{TTF09}, missing values were replaced with the median of the non-missing values for that feature, calculated on the other tasks of the same type performed by the same participant.

\subsection{Classification Settings}
We choose eight popular machine learning classifiers (see Table~\ref{tab:tuning}) based on previous studies using biometrical data~\cite{MF15,KMSL11}.
\input{tables/table_tuning.tex}

In the LORO setting (i.e., the same of the original study), the evaluation on a test set is repeated for 28 times---i.e., the number of participants in our dataset.
At each iteration, we use all observations from
\textit{n-1} participants (i.e., 27) for training the model and test the performance on the remaining one.

In the Hold-out setting, we assess to what extent the trained model can generalize task classification on unseen new data from an hold-out test set not specific to a single participant.
In such a setting, we split the entire dataset into training (20 participants) and test (8 participants) sets.
The model is trained on the entire training set and then evaluated on the held-out test set.
We repeat this process 10 times to further increase the validity of the results.

In line with previous research~\cite{TMH16, TMH18},  in both settings we performed hyper-parameter tuning using the \texttt{caret} R package\footnote{http://topepo.github.io/caret/index.html}.
Table~\ref{tab:tuning} reports the parameters we tuned for to each classifier.
For tuning, we followed a GridSearch approach~\cite{BBB11} with \texttt{tuneLength} = 5---i.e., the maximum number of different values to be evaluated for each parameter~\cite{K08,TMHK17}.

We evaluate the models using precision, recall, and F1-score for which we report values macro-averaged over the evaluation runs~\cite{S02}.
However, for comparison, we focus on the metric reported in the original study---i.e., balanced accuracy (BAC).

%% file: tables/table_features.tex
\begin{table*}[ht]
\caption{Machine learning features grouped by physiological signal.}
\label{tab:features}
\begin{center}
\begin{tabular}{@{}llll@{}}
\toprule
      & \textbf{Signal} & \textbf{Features}            &  \\ 
\midrule                                                                                     
Brain & EEG    & \begin{tabular}[c]{@{}l@{}}
                $\bullet$ Frequency bin for alpha, beta, gamma, delta and theta waves\\ 
                $\bullet$ Ratio between frequency bin of each band and one another \\
                $\bullet$ For the attention and meditation measures: \\ \quad min, max, difference between the mean attention (meditation) during the baseline and during the task\end{tabular} &  \\
\midrule                                                                                     
    \multirow{2}*{Skin} & EDA tonic  & $\bullet$ mean tonic signal     &  \\
                         & EDA phasic & \begin{tabular}[c]{@{}l@{}}
                         $\bullet$ area under the receiving operator curve (AUC) \\ 
                         $\bullet$ min, max, mean, sum peaks amplitudes\end{tabular} &  \\
\midrule                     
 \multirow{3}{*}   {Heart} & BVP   &             
        \begin{tabular}[c]{@{}l@{}}
        $\bullet$ min, max, mean, sum peak amplitudes\\ 
        $\bullet$ difference between the mean peak amplitude during baseline and during the task\end{tabular}   &  \\
       
    & HR    
        & \begin{tabular}[c]{@{}l@{}}$\bullet$ difference between the mean heart rate during the baseline and during the task\\ $\bullet$ difference between the variance heart rate during the baseline and during the task\end{tabular} &  \\
    & \begin{tabular}[c]{@{}l@{}}
    HRV\end{tabular} & 
        \begin{tabular}[c]{@{}l@{}} 
        $\bullet$ standard deviation of beat-to-beat intervals\\ $\bullet$ root mean square of the successive differences \end{tabular} &  \\ 
                              
 \bottomrule
\end{tabular}
\end{center}
\end{table*}

%% file: tables/table_tuning.tex
\begin{table*}[ht]
\caption{Machine learning classifiers used in this study and  parameters used for tuning (``?'' indicates a boolean parameter).}
\label{tab:tuning}
\begin{center}
\begin{tabular}{llll}
\toprule
 \textbf{Family} & \textbf{Classifier (short name)} & \textbf{Parameter} & \textbf{Description} \\ 
 \midrule
      {Bayesian} & {Naive Bayes (nb)} & 
     \begin{tabular}[c]{@{}l@{}} 
             fL\\ 
             usekernel? \\
             adjust
        \end{tabular} &
        \begin{tabular}[c]{@{}l@{}} 
             Laplace correction factor\\ 
             Use kernel density estimate \\
             Bandwidth adjustment
        \end{tabular} \\
 \midrule
 \rowcolor[HTML]{EFEFEF} 
 Nearest Neighbor & K-Nearest Neighbor (knn) & k &\#Clusters \\
 \midrule
 Decision Trees & C4.5-like trees (J48) & C & Confidence factor for pruning \\
 \midrule
 \rowcolor[HTML]{EFEFEF} 
 Support Vector Machines & SVM with Linear Kernel (svmLinear) & C & Cost penalty factor \\
 \midrule
 Neural Networks & Multi-layer Perceptron (mlp) & size & \#Hidden units \\
 \midrule
 \rowcolor[HTML]{EFEFEF} 
 Rule-based & Repeated Incremental Pruning to Produce Error Reduction (Jrip) & NumOpt & \#Optimizations iterations \\
 \midrule
 Bagging & Random Forest (randomforest) & mtry & \#Predictors sampled \\
 \midrule 
 \rowcolor[HTML]{EFEFEF} 
 Boosting & C5.0 & 
     \begin{tabular}[c]{@{}l@{}} 
                 trials\\ 
                 model\\
                 winnow?
        \end{tabular} & 
        \begin{tabular}[c]{@{}l@{}} 
                 \# Boosting iterations\\ 
                 Decision Trees or rule-based\\
                 Apply predictor feature selection 
        \end{tabular}   \\
\bottomrule
\end{tabular}
\end{center}
\end{table*}

%% file: sections/experiment.tex
\begin{figure*}[!h]
    \centering
    \begin{subfigure}[b]{.89\textwidth}
    \includegraphics[width=\linewidth]{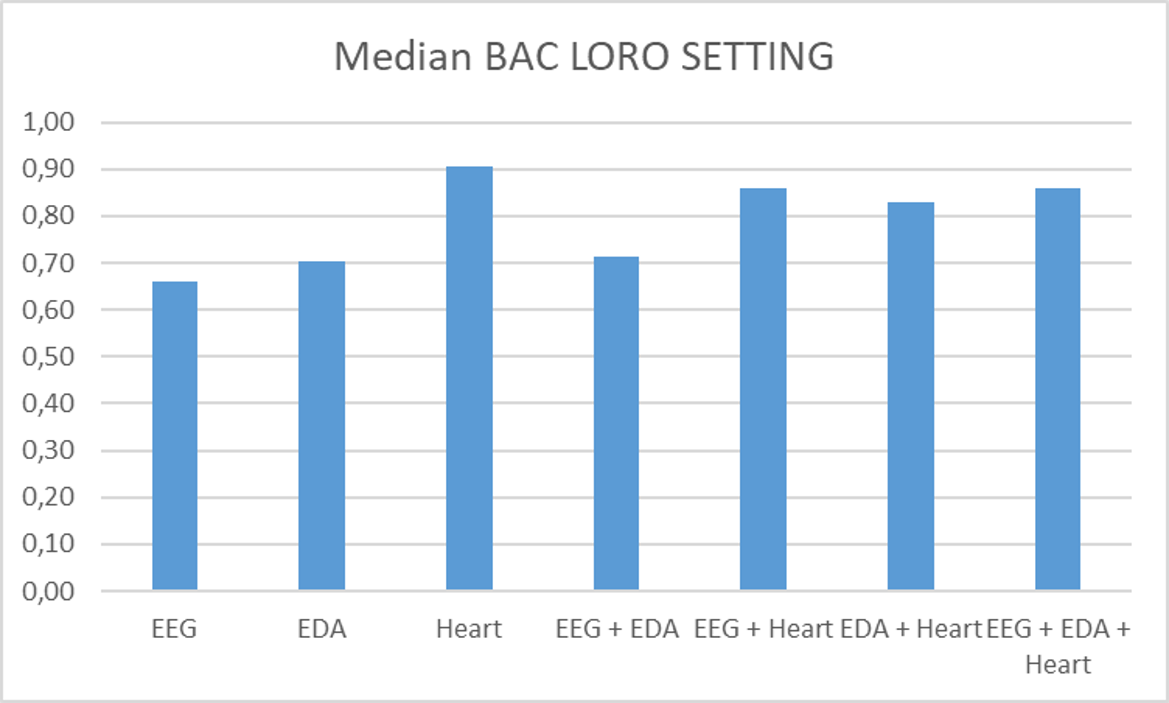}
    \caption{Median BAC of the machine learning classifiers after LORO cross-validation. Results are grouped according to different signal configurations.}
    \label{fig:LORO_bars_chart}
    \end{subfigure}
\begin{subfigure}[b]{.89\textwidth}
     \includegraphics[width=\linewidth] {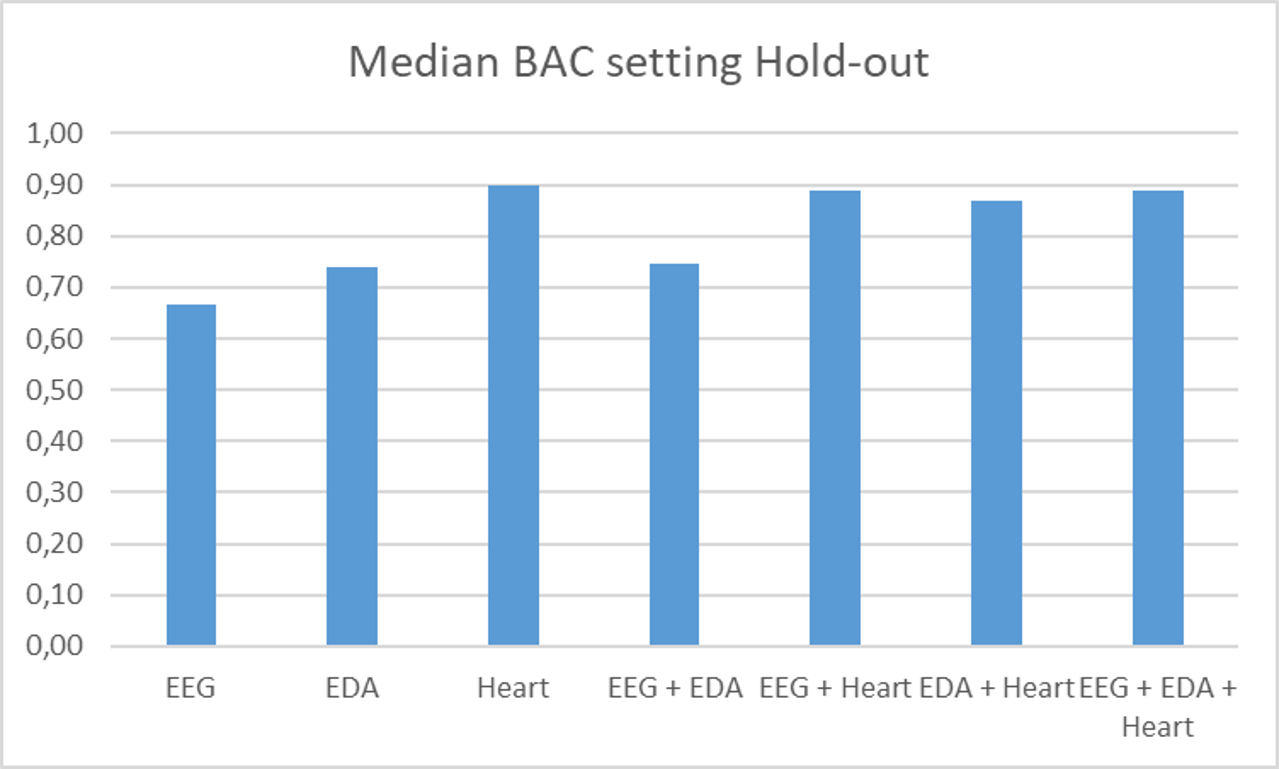}
    \caption{Median BAC of the machine learning classifiers after Hold-out cross-validation. Results are grouped according to different signal configurations.}
    \label{fig:Holdout_bars_chart}
    \end{subfigure}
    \caption{Median BAC of the machine learning classifiers evaluated in this study. Results are grouped according to different signal configurations.}
\end{figure*}

\section{Results}
\label{sec:results}
In this section, we present the results and answer the research questions.

\subsection{$RQ_{Clf}$---Classification of Tasks based on Biometrics}
\label{sec:classification}
We evaluate whether our models can classify which task a participant is performing based on signals combinations from different biometric sensors.
In Figure~\ref{fig:LORO_bars_chart} and Figure~\ref{fig:Holdout_bars_chart}, we compare the median BAC obtained on the different combination of signals by each classifier in the LORO and Hold-out settings respectively.
We did not observe a classifier which performs better than the others independently from the considered signal(s).
On the other hand, \textit{nb} and \textit{knn} classifiers do not seem appropriate for our classification task.

In Table~\ref{tab:best_results_LORO} and Table~\ref{tab:best_results_HoldOut}, for each signal and their combination, we report the classifier with the highest BAC together with its precision, recall, and F-measure.
In both evaluation settings, the EEG signal shows the worst performance (BAC = 66\%, 67\%).
Conversely, Heart is the signal with the highest performance, yielding a BAC of 87\% (90\% in the Hold-out evaluation setting).
Moreover, when combined with Hearth, EEG and EDA performances increase considerably.
This suggests that the best and most reliable classification results can be accomplished solely using the Empatica E4 sensors while the EEG contribution is negligible.

We answer $RQ_{Clf}$ as follows.
\begin{mdframed}[backgroundcolor=lightgray!20]
Physiological signals can be used to train classifiers which accurately differentiate between code and prose comprehension tasks. The classifier trained using features based on Heart signal shows the best results.
\end{mdframed}

\begin{table}[ht]
\caption{Results of the best machine learning classifier evaluated using LORO cross-validation.}
\label{tab:best_results_LORO}
\begin{minipage}{\columnwidth}
\begin{center}
\begin{tabular}{lllllll}
\toprule
\textbf{Signal} & \shortstack{\textbf{Best}\\\textbf{Classifier}} & \textbf{Precision} & \textbf{Recall} & \textbf{F1} & \textbf{BAC} \\ \midrule
EEG & mlp & 0.72 & 0.66 & 0.62 & 0.66 \\
\rowcolor[HTML]{EFEFEF}
EDA & rf & 0.78 & 0.71 & 0.71 & 0.71 \\

Heart & mlp & 0.91 & 0.87 & 0.87 & \textbf{0.87} \\
\rowcolor[HTML]{EFEFEF}
EEG + EDA & C5.0 & 0.75 & 0.72 & 0.72 & 0.72 \\

EEG + Heart & Jrip & 0.90 & 0.86 & 0.87 & 0.86 \\
\rowcolor[HTML]{EFEFEF}
EDA + Heart & mlp & 0.91 & 0.83 & 0.86 & 0.83 \\

EEG + EDA + Heart & Jrip & 0.88 & 0.86 & 0.86 & 0.86 \\
\bottomrule
\end{tabular}

\end{center}
\end{minipage}
\end{table}

\begin{table}[ht]
\caption{Results of the best machine learning classifier evaluated using Hold-out cross-validation.}
\label{tab:best_results_HoldOut}
\begin{minipage}{\columnwidth}
\begin{center}
\begin{tabular}{lllllll}
\toprule
\textbf{Signal} & \shortstack{\textbf{Best}\\\textbf{Classifier}} & \textbf{Precision} & \textbf{Recall} & \textbf{F1} & \textbf{BAC} \\ \midrule
EEG & rf & 0.70 & 0.67 & 0.68 & 0.67 \\
\rowcolor[HTML]{EFEFEF}
EDA & Knn & 0.83 & 0.74 & 0.77 & 0.74 \\
Heart & mlp & 0.95 & 0.90 & 0.92 & \textbf{0.90} \\\rowcolor[HTML]{EFEFEF}
EEG + EDA & mlp & 0.75 & 0.75 & 0.75 & 0.75 \\
EEG + Heart & C5.0 & 0.90 & 0.89 & 0.90 & 0.89 \\\rowcolor[HTML]{EFEFEF}
EDA + Heart & svm & 0.93 & 0.87 & 0.89 & 0.87  \\
EEG + EDA + Heart & C5.0 & 0.92 & 0.89 & 0.90 & 0.89 \\
\bottomrule
\end{tabular}
\end{center}
\end{minipage}
\end{table}

\subsection{$RQ_{Exp}$---Classification Accuracy and Expertise}
\label{sec:expertise}
We examine the relationship between classifier accuracy and participant expertise.
For each participant, we consider the classifier with the best BAC among all configurations of classifiers and signals.
We then calculate the association between BAC and the participant's GPA using the Kendall tau correlation coefficient ($\tau \in [-1, -1]$ with 0 indicating no association).
For statistical testing ($\alpha$ = 0.05), the null hypothesis assumes no association between the two variables.

\begin{figure}[!htb]
    \centering
    \includegraphics[width=.95\linewidth]{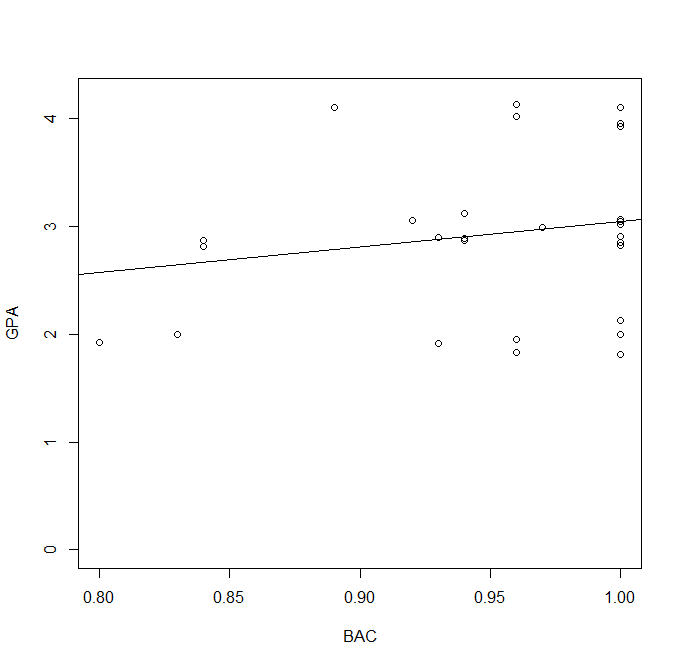}
    \caption{Scatterplot of classifiers BAC and participants' GPA. The regression line indicates a non-significant relationship.}
    \label{fig:experitse}
\end{figure}

There is a positive, although small, correlation between the classifier accuracy and the participants' expertise ($\tau$ = 0.16), as shown in Fig. \ref{fig:experitse}.
However, we failed to reject the null hypothesis ($p$ = 0.31).
We answer $RQ_{Exp}$ as follows.
\begin{mdframed}[backgroundcolor=lightgray!20]
Expertise is not related to the accuracy of classifiers trained using biometrical signals.
\end{mdframed}

\subsection{Comparison with results of the original study}
$RQ_{Clf}$ \textit{replicates} the results of RQ1 from the original study.
For comparison, we contrast our settings (code comprehension vs. prose comprehension) with the ones reported in the original study (code comprehension vs. prose review).
We consider the best configuration of classifier and signal---i.e., \texttt{mlp} and Heart.
Table~\ref{tab:result_comparison} reports the best classifier BAC for both classes.
\input{tables/table_comparisonResults.tex}

In both studies, prose-related tasks are better identified than code-related ones and with high accuracy---$95\%$ in the original study and $99\%$ in the replication.
Our approach improves by 8\% (9\% if considering the Hold-out evaluation setting) the original study accuracy when classifying the same code comprehension tasks.
Finally, the overall results show an improvement of 8\% (11\% if considering the Hold-out evaluation setting) obtained by using lightweight biometric sensors rather than fMRI.

We did \textit{not replicate} $RQ_{Exp}$---i.e., RQ3 in the original study (see Table~\ref{tab:result_comparison}).
The original study reported a negative correlation coefficient (Pearson $r$ = -0.44, $p$ = 0.016) between classifier accuracy and expertise.
The inverse relationship between accuracy and expertise suggests that neural representations of code and prose are harder to differentiate once coding skills increase (and viceversa).
Given the non-significant and small correlation coefficient reported in this study, a similar relationship between expertise and heart-related signals is not apparent.
We believe that the reason for this result is the limited variation of experience within our sample.
Compared to the original study (i.e., a mix of graduate and undergraduate students from different schools), our sample is more homogeneous (i.e., undergraduate students from a single school) which can be reflected in their expertise.

%% file: tables/table_comparisonResults.tex
        

\begin{table}[!ht]
\centering
\caption{Result comparison between the original study and this replication. For $RQ_{Clf}$, best BAC results for LORO (Hold-out) validation are reported.}
\label{tab:result_comparison}
\scriptsize
\begin{tabular}{clccc}
\toprule
 \textbf{\begin{tabular}[c]{@{}c@{}}RQ\\(original study) \end{tabular}}&  & \multicolumn{1}{c}{\textbf{\begin{tabular}[c]{@{}c@{}}Original \\ Study\end{tabular}}} & \multicolumn{1}{c}{\textbf{\begin{tabular}[c]{@{}c@{}}This \\ replication\end{tabular}}} & \multicolumn{1}{c}{\textbf{Replicated?}} \\ \midrule
\multirow{3}{*}{\begin{tabular}[c]{@{}l@{}}$RQ_{Clf}$\\ (RQ1)\end{tabular}} & \begin{tabular}[c]{@{}l@{}}Overall\end{tabular} & 0.79 & 0.87 (0.90) & \multirow{3}{*}{\begin{tabular}[c]{@{}c@{}}Yes,\\with improvements\end{tabular}} \\
 & \begin{tabular}[c]{@{}l@{}} Code\end{tabular} & 0.72 & 0.80 (0.81) &  \\
 & \begin{tabular}[c]{@{}l@{}} Prose\end{tabular} & 0.95 & 0.99 (0.99) &  \\ \midrule
\begin{tabular}[c]{@{}c@{}}$RQ_{Exp}$\\(RQ3)\end{tabular} & & \begin{tabular}[c]{@{}c@{}}$r$ = -0.44 \\($p$ = 0.01) \end{tabular} &  \begin{tabular}[c]{@{}c@{}}$\tau$ = 0.16\\ ($p$ = 0.31)  \end{tabular} & No \\ \bottomrule
\end{tabular}%
\end{table}

%% file: sections/discussion.tex
\section{Discussion}
\label{sec:discussion}
In this section, we discuss the limitations of our study, the current state of knowledge based on this and the original study results, and consider implications for practice and research. 

\subsection{Threats to Validity}
Threats to the external validity of our replication are associated with the representativeness of the tasks.
For code comprehension, we used the same tasks of the original study which, although targeting only the C language, were sampled from real-world projects.
For prose comprehension, we use standardized text from the Italian Ministry of Education. 

Our study suffers from threats to construct validity---i.e., the reliability of are our measures in capturing code and prose comprehension.
For the former, we use the same questions as in the original study~\cite{FSW17}, which are also utilized in previous work (e.g.,~\cite{SMD06}).
For the latter, we use examples from cognitive linguistic studies~\cite{HSG02,Mac06}.
The complex biometric signals needed to be filtered before the analysis. 
To that end, we followed state-of-the-art practices from signal processing.
As in the original study, the expertise construct is measured through a proxy---i.e., GPA.
Although we acknowledge that software development expertise is difficult to measure,  GPA correlates with learning and academic skills~\cite{GWG06}, and was measured taking into only courses focusing on the programming language used during the experiment.

When assessing the impact of expertise on classifiers performance, we did not observe a significant correlation. 
We cannot exclude that such result is due to the homogeneity of our sample which includes only undergraduate students; thus, representing a potential threat to internal validity.

The validity of our conclusions is based on the robustness of the machine learning models and null-hypothesis statistical test used to answer the research questions. 
Regarding machine learning, we mitigated such threat by i) running several algorithms addressing the same classification task, ii) applying hyper-parameters tuning to optimally solve the task, and iii) reporting results from two different evaluation settings---i.e., LORO and Hold-out.
Regarding hypothesis testing, we rely on robust, non-parametric statistics~\cite{Noe81} for which the effect size (i.e., Kendall $\tau$) can be interpreted similarly to the one reported in the original study (i.e., Pearson $r$).

\subsection{Drawing Conclusion Across Studies}
In this study, we strengthen the original study conclusion that different comprehension tasks can be recognized using biometric-based signals as a proxy for cognitive effort.
In particular, we explicitly compare code comprehension and prose comprehension---strengthening the construct validity of the original study---while demonstrating an affordable approach in real-world settings by using lightweight sensors.
Our setup costs less than \$2,000 as opposed to the \$21,000 reported in the original study.
Considering performance on individual tasks, we show that prose comprehension can be more accurately (and in almost every case is) recognized compared to code comprehension.  
Our best overall accuracy (BAC = 90\%) and the improvement over the original study ($\Delta$BAC = 8\%, approx. \$2,400 saved for each percentage point gained in performance) is ground for further evaluation in in-vivo settings---e.g., integration with development environments.

\textbf{Implications for Practice}. 
Our results provide opportunities for improving software engineering tools. 
By means of a relatively cheap wristband (approx. \$1,500), the development environment can become aware of the comprehension task in which a developer is engaged and optimize support for such a task. 
For example, knowing when developers are comprehending source code---a more demanding task than prose comprehension~\cite{FBM14}---can be leveraged to better measure their interruptibility~\cite{ZCM17} and adjust their environment (e.g., by temporarily disabling notifications). 

Using our approach, developers can collect interesting metrics from a Personal Software Process perspective~\cite{Hum02}, such as time spent comprehending source code vs. time spent understanding requirement specifications or documentation. 
Developers can leverage these metrics to improve their productivity, effort estimation, and planning skills.

Furthermore, a developer engaged for too long in comprehending a specification or a piece of code can indicate quality issues related to complexity~\cite{YKY16}. 
In the future, the integration of our approach with eye-tracking will allow us to identify the specific focus of a developer~\cite{KWS17,SSW16} and recommend  documentation to support her information needs~\cite{RMT17, ECN19}. 

\textbf{Implications for Research}. 
Although previous studies have tackled the use of lightweight biometrics in software engineering, this study is one of the first to explicitly deal with code comprehension. 

Currently, researchers study comprehension strategies (e.g., bottom-up vs. top-down) by relying on developers' assessment (e.g., subjective rating~\cite{DR00} or think aloud protocols~\cite{SV95}) or more invasive methods (e.g., fMRI~\cite{SKA14}).
The ability to automatically recognize code comprehension tasks using physiological signals enables less invasive research on comprehension strategies. 
Code comprehension is the basis for several other software development tasks~\cite{FSW17}. 
Our approach can be used to study the ``weight'' that comprehension has for tasks such as refactoring~\cite{FSP12} and code reviews~\cite{Van01} in an unobtrusive (and cheaper) way.

Cognitive activities are related to task difficulty. 
As shown in the original study~\cite{FSW17}, understanding code is more difficult than comprehending text. 
Our study confirms previous work results which showed that it is possible to classify task difficulty using lightweight biometric sensors~\cite{FBM14}.

We showed that an EEG headset equipped with one electrode is not sufficient to recognize the task a participant is performing.
Therefore, we suggest to researchers interested in the same goal of this study, but focusing on the investigation of neural activity measured unobtrusively through EEG, to invest in devices with higher definition (e.g., 14 or 32 channels).

Furthermore, researchers can replicate our setup using devices available at retail shops and standard data analysis tools.

%% file: sections/conclusion.tex
\section{Conclusion and Future Work}
\label{sec:conclusion}
This paper presents the replication of a previous study aimed at i) automatically classifying which kind of comprehension task (prose or code) a developer is performing and ii) studying the correlation between classifier accuracy and expertise. 

In the original study, the authors explored the use of fMRI finding that it is possible to classify which task a participant is undertaking based on brain activity. 
However, collecting fMRI signals is expensive and can be applied only for \textit{in-vitro} studies. 
Therefore, we investigated whether fMRI could be replaced by lightweight devices, which previous research used to investigate cognitive effort in software development.
We found that an off-the-shelf EEG headset is not suitable to achieve our goal with high performance. 
Conversely, the heart activity, captured using a wristband, can be used to distinguish between code and prose comprehension tasks with high accuracy.  
The original study also showed that, when considering expert developers, the two tasks are harder to distinguish at neural level.
We were not able to replicate this result using biometric signals in our homogeneous sample composed of undergraduate students. 
Further replications, involving a more heterogeneous sample, are required to further investigate the association between participants' expertise and the performance of the task classifiers. 

Our future work consists in investigating software development expertise from a physiological perspective.
Furthermore, we want to assess an additional task in which code and prose are mixed such as technical documentation, programming tutorials, and StackOverflow posts.

%% file: main.bbl
\begin{thebibliography}{10}
\providecommand{\url}[1]{#1}
\csname url@samestyle\endcsname
\providecommand{\newblock}{\relax}
\providecommand{\bibinfo}[2]{#2}
\providecommand{\BIBentrySTDinterwordspacing}{\spaceskip=0pt\relax}
\providecommand{\BIBentryALTinterwordstretchfactor}{4}
\providecommand{\BIBentryALTinterwordspacing}{\spaceskip=\fontdimen2\font plus
\BIBentryALTinterwordstretchfactor\fontdimen3\font minus
  \fontdimen4\font\relax}
\providecommand{\BIBforeignlanguage}[2]{{%
\expandafter\ifx\csname l@#1\endcsname\relax
\typeout{** WARNING: IEEEtran.bst: No hyphenation pattern has been}%
\typeout{** loaded for the language `#1'. Using the pattern for}%
\typeout{** the default language instead.}%
\else
\language=\csname l@#1\endcsname
\fi
#2}}
\providecommand{\BIBdecl}{\relax}
\BIBdecl

\bibitem{CLB03}
M.~Ciolkowski, O.~Laitenberger, and S.~Biffl, ``{Software Reviews: The State of
  the Practice},'' \emph{IEEE software}, no.~6, pp. 46--51, 2003.

\bibitem{LMC04}
F.~Lanubile, T.~Mallardo, F.~Calefato, C.~Denger, and M.~Ciolkowski,
  ``{Assessing the Impact of Active Guidance for Defect Detection: A Replicated
  Experiment},'' in \emph{null}.\hskip 1em plus 0.5em minus 0.4em\relax IEEE,
  2004, pp. 269--279.

\bibitem{RCJ08}
D.~Rombach, M.~Ciolkowski, R.~Jeffery, O.~Laitenberger, F.~McGarry, and
  F.~Shull, ``{Impact of Research on Practice in the Field of Inspections,
  Reviews and Walkthroughs: Learning from Successful Industrial Uses},''
  \emph{ACM SIGSOFT Software Engineering Notes}, vol.~33, no.~6, pp. 26--35,
  2008.

\bibitem{NMV11}
K.~Nishizono, S.~Morisakl, R.~Vivanco, and K.~Matsumoto, ``{Source Code
  Comprehension Strategies and Metrics to Predict Comprehension Effort in
  Software Maintenance and Evolution Tasks-An Empirical Study with Industry
  Practitioners},'' in \emph{Software Maintenance (ICSM), 2011 27th IEEE
  International Conference on}.\hskip 1em plus 0.5em minus 0.4em\relax IEEE,
  2011, pp. 473--481.

\bibitem{ARM18}
A.~Armaly, P.~Rodeghero, and C.~McMillan, ``{AudioHighlight: Code Skimming for
  Blind Programmers},'' in \emph{2018 IEEE International Conference on Software
  Maintenance and Evolution (ICSME)}.\hskip 1em plus 0.5em minus 0.4em\relax
  IEEE, 2018, pp. 206--216.

\bibitem{SKA14}
J.~Siegmund, C.~K{\"a}stner, S.~Apel, C.~Parnin, A.~Bethmann, T.~Leich,
  G.~Saake, and A.~Brechmann, ``{Understanding Understanding Source Code with
  Functional Magnetic Resonance Imaging},'' in \emph{Proceedings of the 36th
  International Conference on Software Engineering}.\hskip 1em plus 0.5em minus
  0.4em\relax ACM, 2014, pp. 378--389.

\bibitem{FSW17}
B.~Floyd, T.~Santander, and W.~Weimer, ``{Decoding the Representation of Code
  in the Brain: An fMRI Study of Code Review and Expertise},'' in
  \emph{Proceedings of the 39th International Conference on Software
  Engineering}.\hskip 1em plus 0.5em minus 0.4em\relax IEEE Press, 2017, pp.
  175--186.

\bibitem{ECN19}
F.~Ebert, F.~Castor, N.~Novielli, and A.~Serebrenik, ``{Confusion in Code
  Reviews: Reasons, Impacts, and Coping Strategies},'' in \emph{Proceedings of
  26th International Conference on Software Analysis, Evolution and
  Reengineering {SANER 2019}}.\hskip 1em plus 0.5em minus 0.4em\relax IEEE
  Press, 2019, pp. 49--60.

\bibitem{ZCM17}
M.~Z{\"u}ger, C.~Corley, A.~N. Meyer, B.~Li, T.~Fritz, D.~Shepherd,
  V.~Augustine, P.~Francis, N.~Kraft, and W.~Snipes, ``{Reducing Interruptions
  at Work: A Large-scale Field Study of FlowLight},'' in \emph{Proceedings of
  the 2017 CHI Conference on Human Factors in Computing Systems}.\hskip 1em
  plus 0.5em minus 0.4em\relax ACM, 2017, pp. 61--72.

\bibitem{FBM14}
T.~Fritz, A.~Begel, S.~C. M{\"u}ller, S.~Yigit-Elliott, and M.~Z{\"u}ger,
  ``{Using Psycho-physiological Measures to Assess Task Difficulty in Software
  Development},'' in \emph{Proceedings of the 36th International Conference on
  Software Engineering}.\hskip 1em plus 0.5em minus 0.4em\relax ACM, 2014, pp.
  402--413.

\bibitem{FMA18}
S.~Fakhoury, Y.~Ma, V.~Arnaoudova, and O.~Adesope, ``{The Effect of Poor Source
  Code Lexicon and Readability on Developers' Cognitive Load},'' in \emph{Proc.
  Int'l Conf. Program Comprehension (ICPC)}, 2018.

\bibitem{RSI98}
D.~W. Rowe, J.~Sibert, and D.~Irwin, ``{Heart Rate Variability: Indicator of
  User State as an Aid to Human-computer Interaction},'' in \emph{Proceedings
  of the SIGCHI conference on Human factors in computing systems}.\hskip 1em
  plus 0.5em minus 0.4em\relax ACM Press/Addison-Wesley Publishing Co., 1998,
  pp. 480--487.

\bibitem{GLP18}
M.~Gjoreski, M.~Lu{\v{s}}trek, and V.~Pejovi{\'c}, ``{My Watch Says I'm Busy:
  Inferring Cognitive Load with Low-Cost Wearables},'' in \emph{Proceedings of
  the 2018 ACM International Joint Conference and 2018 International Symposium
  on Pervasive and Ubiquitous Computing and Wearable Computers}.\hskip 1em plus
  0.5em minus 0.4em\relax ACM, 2018, pp. 1234--1240.

\bibitem{Car10}
J.~C. Carver, ``{Towards Reporting Guidelines for Experimental Replications: A
  Proposal},'' in \emph{1st International Workshop on Replication in Empirical
  Software Engineering Research}, 2010.

\bibitem{BCD14}
M.~T. Baldassarre, J.~Carver, O.~Dieste, and N.~Juristo, ``{Replication Types:
  Towards a Shared Taxonomy},'' in \emph{Proceedings of the 18th International
  Conference on Evaluation and Assessment in Software Engineering}.\hskip 1em
  plus 0.5em minus 0.4em\relax ACM, 2014, p.~18.

\bibitem{SCV08}
F.~J. Shull, J.~C. Carver, S.~Vegas, and N.~Juristo, ``{The Role of
  Replications in Empirical Software Engineering},'' \emph{Empirical software
  engineering}, vol.~13, no.~2, pp. 211--218, 2008.

\bibitem{PSA18}
N.~Peitek, J.~Siegmund, S.~Apel, C.~K{\"a}stner, C.~Parnin, A.~Bethmann,
  T.~Leich, G.~Saake, and A.~Brechmann, ``{A Look into Programmers' Heads},''
  \emph{IEEE Transactions on Software Engineering}, 2018.

\bibitem{RHM15}
S.~Radevski, H.~Hata, and K.~Matsumoto, ``{Real-time Monitoring of Neural State
  in Assessing and Improving Software Developers' Productivity},'' in
  \emph{Proceedings of the Eighth International Workshop on Cooperative and
  Human Aspects of Software Engineering}.\hskip 1em plus 0.5em minus
  0.4em\relax IEEE Press, 2015, pp. 93--96.

\bibitem{MF15}
S.~M{\"u}ller and T.~Fritz, ``{Stuck and Frustrated or in Flow and Happy:
  Sensing Developers' Emotions and Progress},'' in \emph{Software Engineering
  (ICSE), 2015 IEEE/ACM 37th IEEE International Conference on}, vol.~1.\hskip
  1em plus 0.5em minus 0.4em\relax IEEE, 2015, pp. 688--699.

\bibitem{Par11}
C.~Parnin, ``{Subvocalization-Toward Hearing the Inner Thoughts of
  Developers},'' in \emph{Program Comprehension (ICPC), 2011 IEEE 19th
  International Conference on}.\hskip 1em plus 0.5em minus 0.4em\relax IEEE,
  2011, pp. 197--200.

\bibitem{MF16}
S.~M{\"u}ller and T.~Fritz, ``{Using (bio) Metrics to Predict Code Quality
  Online},'' in \emph{Proceedings of the 38th International Conference on
  Software Engineering}.\hskip 1em plus 0.5em minus 0.4em\relax ACM, 2016, pp.
  452--463.

\bibitem{IU14}
Y.~Ikutani and H.~Uwano, ``{Brain Activity Measurement During Program
  Comprehension with NIRS},'' in \emph{Software Engineering, Artificial
  Intelligence, Networking and Parallel/Distributed Computing (SNPD), 2014 15th
  IEEE/ACIS International Conference on}.\hskip 1em plus 0.5em minus
  0.4em\relax IEEE, 2014, pp. 1--6.

\bibitem{PSP18}
N.~Peitek, J.~Siegmund, C.~Parnin, S.~Apel, J.~Hofmeister, and A.~Brechmann,
  ``{Simultaneous Measurement of Program Comprehension with fMRI and Eye
  Tracking: A Case Study},'' in \emph{Proc. Int’l Symposium Empirical
  Software Engineering and Measurement (ESEM)}.\hskip 1em plus 0.5em minus
  0.4em\relax ACM, 2018.

\bibitem{BWJR15}
J.~J. Braithwaite, D.~G. Watson, R.~Jones, and M.~Rowe, ``{A Guide for
  Analysing Electrodermal Activity (EDA) \& Skin Conductance Responses (SCRs)
  for Psychological Experiments},'' \emph{Psychophysiology}, vol.~49, no.~1,
  pp. 1017--1034, 2013.

\bibitem{CFSGL11}
F.~Canento, A.Fred, H.~Silva, H.~Gamboa, and A.~Lourenço, ``Multimodal
  biosignal sensor data handling for emotion recognition,'' in \emph{SENSORS,
  2011 IEEE}, 2011, pp. 647--650.

\bibitem{GVLSC16}
A.~Greco, G.~Valenza, A.~Lanata, E.~P. Scilingo, and L.~Citi, ``{cvxEDA: A
  Convex Optimization Approach to Electrodermal Activity Processing},''
  \emph{IEEE Transactions on Biomedical Engineering}, vol.~63, no.~4, pp.
  797--804, 2016.

\bibitem{TTF09}
H.~Trevor, R.~Tibshirani, and J.~Friedman, \emph{The Elements of Statistical
  Learning: Data Mining, Inference, and Prediction}, 2nd~ed.\hskip 1em plus
  0.5em minus 0.4em\relax Springer, 2009.

\bibitem{KMSL11}
S.~Koelstra, C.~M\"uhl, M.~Soleymani, J.-S. Lee, A.~Yazdani, T.~Ebrahimi,
  T.~Pun, A.~Nijholt, and I.~Yiannis)~Patras, ``Deap: A database for emotion
  analysis using physiological signals,'' \emph{IEEE Transactions on Affective
  Computing}, vol.~3, pp. 18--31, 12 2011.

\bibitem{TMH16}
C.~Tantithamthavorn, S.~McIntosh, A.~E. Hassan, and K.~Matsumoto, ``Automated
  parameter optimization of classification techniques for defect prediction
  models,'' in \emph{2016 IEEE/ACM 38th International Conference on Software
  Engineering (ICSE)}.\hskip 1em plus 0.5em minus 0.4em\relax IEEE, 2016, pp.
  321--332.

\bibitem{TMH18}
------, ``The impact of automated parameter optimization on defect prediction
  models,'' \emph{IEEE Transactions on Software Engineering}, 2018.

\bibitem{BBB11}
J.~S. Bergstra, R.~Bardenet, Y.~Bengio, and B.~K{\'e}gl, ``{Algorithms for
  Hyper-parameter Optimization},'' in \emph{Advances in neural information
  processing systems}, 2011, pp. 2546--2554.

\bibitem{K08}
M.~Kuhn, ``{Building Predictive Models in R Using the caret Package},''
  \emph{Journal of Statistical Software, Articles}, vol.~28, no.~5, pp. 1--26,
  2008.

\bibitem{TMHK17}
C.~Tantithamthavorn, S.~McIntosh, A.~E. Hassan, and K.~Matsumoto, ``{An
  Empirical Comparison of Model Validation Techniques for Defect Prediction
  Models},'' \emph{IEEE Transactions on Software Engineering}, vol.~43, no.~1,
  pp. 1--18, 2017.

\bibitem{S02}
F.~Sebastiani, ``Machine learning in automated text categorization,'' \emph{ACM
  Comput. Surv.}, vol.~34, no.~1, pp. 1--47, Mar. 2002.

\bibitem{SMD06}
J.~Sillito, G.~C. Murphy, and K.~De~Volder, ``{Questions Programmers ask During
  Software Evolution Tasks},'' in \emph{Proceedings of the 14th ACM SIGSOFT
  international symposium on Foundations of software engineering}.\hskip 1em
  plus 0.5em minus 0.4em\relax ACM, 2006, pp. 23--34.

\bibitem{HSG02}
J.~Hatcher, M.~J. Snowling, and Y.~M. Griffiths, ``{Cognitive Assessment of
  Dyslexic Students in Higher Education},'' \emph{British journal of
  educational psychology}, vol.~72, no.~1, pp. 119--133, 2002.

\bibitem{Mac06}
E.~Macaro, ``{Strategies for Language Learning and For Language Use: Revising
  the Theoretical Framework},'' \emph{The Modern Language Journal}, vol.~90,
  no.~3, pp. 320--337, 2006.

\bibitem{GWG06}
W.~A. Grove, T.~Wasserman, and A.~Grodner, ``{Choosing a Proxy for Academic
  Aptitude},'' \emph{The Journal of Economic Education}, vol.~37, no.~2, pp.
  131--147, 2006.

\bibitem{Noe81}
G.~E. Noether, ``{Why Kendall Tau?}'' \emph{Teaching Statistics}, vol.~3,
  no.~2, pp. 41--43, 1981.

\bibitem{Hum02}
W.~S. Humphrey, ``{Personal Software Process (PSP)},'' \emph{Encyclopedia of
  Software Engineering}, 2002.

\bibitem{YKY16}
X.~Yang, R.~G. Kula, N.~Yoshida, and H.~Iida, ``{Mining the Modern Code Review
  Repositories: A Dataset of People, Process and Product},'' in
  \emph{Proceedings of the 13th International Conference on Mining Software
  Repositories}.\hskip 1em plus 0.5em minus 0.4em\relax ACM, 2016, pp.
  460--463.

\bibitem{KWS17}
K.~Kevic, B.~Walters, T.~Shaffer, B.~Sharif, D.~C. Shepherd, and T.~Fritz,
  ``{Eye Gaze and Interaction Contexts for Change Tasks-Observations and
  Potential},'' \emph{Journal of Systems and Software}, vol. 128, pp. 252--266,
  2017.

\bibitem{SSW16}
B.~Sharif, T.~Shaffer, J.~Wise, and J.~I. Maletic, ``{Tracking Developers' Eyes
  in the IDE},'' \emph{IEEE Software}, vol.~33, no.~3, pp. 105--108, 2016.

\bibitem{RMT17}
M.~P. Robillard, A.~Marcus, C.~Treude, G.~Bavota, O.~Chaparro, N.~Ernst, M.~A.
  Gerosa, M.~Godfrey, M.~Lanza, and M.~Linares-V{\'a}squez, ``{On-demand
  Developer Documentation},'' in \emph{Software Maintenance and Evolution
  (ICSME), 2017 IEEE International Conference on}.\hskip 1em plus 0.5em minus
  0.4em\relax IEEE, 2017, pp. 479--483.

\bibitem{DR00}
A.~Dunsmore and M.~Roper, ``{A Comparative Evaluation of Program Comprehension
  Measures},'' \emph{The Journal of Systems and Software}, vol.~52, no.~3, pp.
  121--129, 2000.

\bibitem{SV95}
T.~M. Shaft and I.~Vessey, ``{The Relevance of Application Domain Knowledge:
  The Case of Computer Program Comprehension},'' \emph{Information systems
  research}, vol.~6, no.~3, pp. 286--299, 1995.

\bibitem{FSP12}
J.~Feigenspan, M.~Schulze, M.~Papendieck, C.~K{\"a}stner, R.~Dachselt,
  V.~K{\"o}ppen, M.~Frisch, and G.~Saake, ``{Supporting Program Comprehension
  in Large Preprocessor-based Software Product Lines},'' \emph{IET software},
  vol.~6, no.~6, pp. 488--501, 2012.

\bibitem{Van01}
A.~Van~Deursen, ``{Program Comprehension Risks and Opportunities in Extreme
  Programming},'' in \emph{Reverse Engineering, 2001. Proceedings. Eighth
  Working Conference on}.\hskip 1em plus 0.5em minus 0.4em\relax IEEE, 2001,
  pp. 176--185.

\end{thebibliography}
